\documentclass[prd,twocolumn,showpacs,nofootinbib,amsmath,amssymb]{revtex4-1}
\usepackage{graphicx,dcolumn,bm}
\usepackage{hyperref}
\usepackage{color}
\begin{document}
\title{Indirect and direct detection prospect for TeV dark matter in the MSSM-9}
\author{Maria Eugenia Cabrera-Catalan}
\affiliation{Instituto de F\'{i}sica, Universidade de S\~{a}o Paulo, C.P. 66.318, 05315-970 S\~ao Paulo, Brazil} 
\affiliation{Instituto de F\'{i}sica Te\'orica, IFT-UAM/CSIC, U.A.M. Cantoblanco, 28049 Madrid, Spain}
\affiliation{GRAPPA Institute, University of Amsterdam, 1098 XH Amsterdam, The Netherlands}
\author{Shin'ichiro Ando}
\author{Christoph Weniger}
\author{Fabio Zandanel}
\affiliation{GRAPPA Institute, University of Amsterdam, 1098 XH Amsterdam, The Netherlands}
\date{\today}

\newcommand\ev[1]{\left\langle#1\right\rangle}

\begin{abstract}
  We investigate the prospects of indirect and direct dark matter searches
  within the minimal supersymmetric standard model with nine parameters
  (MSSM-9). These nine parameters include three gaugino masses, Higgs, slepton
  and squark masses, all treated independently. We perform a Bayesian Monte
  Carlo scan of the parameter space taking into consideration all available
  particle physics constraints such as the Higgs mass of 126~GeV, upper limits
  on the scattering cross-section from direct-detection experiments, and
  assuming that the MSSM-9 provides all the dark matter abundance through
  thermal freeze-out mechanism.  Within this framework we find two most
  probable regions for dark matter: 1-TeV higgsino-like and 3-TeV wino-like
  neutralinos. We discuss prospects for future indirect (in particular the
  Cherenkov Telescope Array, CTA) and direct detection experiments.  We find
  that for slightly contracted dark matter profiles in our Galaxy, which can be
  caused by the effects of baryonic infall in the Galactic center, CTA will be
  able to probe a large fraction of the remaining allowed region in synergy
  with future direct detection experiments like XENON-1T.
\end{abstract}

\maketitle

\section{Introduction}
\label{sec:intro}

Identifying the particle nature of dark matter is one of the most pressing
goals of modern astrophysics and cosmology.  If dark matter is made of weakly
interacting massive particles (WIMPs)~\cite{freezeout1, Bertone:2004pz}, where
the relic dark matter abundance is naturally explained with the thermal
freeze-out mechanism, a particular promising avenue is the search for
signatures from the self-annihilation of dark matter particles in gamma
rays~\cite{Bringmann:2012ez}.  The required average velocity-weighted
annihilation cross-section during freeze-out is of the order of $\langle \sigma
v \rangle = 3 \times 10^{-26}$~cm$^{3}$~s$^{-1}$~\cite{freezeout1, freezeout2}.
Upper limits from modern gamma-ray instruments, such as the Large Area
Telescope (LAT) aboard \emph{Fermi} satellite, start to exclude this canonical
annihilation cross-section for WIMP masses below 100~GeV~\cite{FermiDwarf}. At
the same time, the hunt for signatures of physics beyond the standard model of
particle physics, especially of supersymmetry, is ongoing at the Large Hadron
Collider (LHC).  Supersymmetric models provide interesting candidate particles
for WIMP dark matter (in most cases the lightest neutralino). Although there
has been no claim of positive signatures of supersymmetry yet, some outstanding
conclusions can be drawn from several measurements at the LHC. One of them
comes from the mass measurement of the Higgs boson of about
126~GeV~\cite{LHC-Higgs1, LHC-Higgs2}.  

Since particle masses and couplings are subject to loop corrections, the Higgs
mass measurement can provide indirect information about yet undiscovered
particles. Therefore, this measurement already significantly constrains the
parameter space of supersymmetric models.  Reference~\cite{Cabrera:2012}
studied the implications of the Higgs mass measurement for the minimal
supersymmetric standard model (MSSM) with five parameters (the constrained
MSSM), and found that the posterior probabilities for these parameters are
narrowly distributed.  Reference~\cite{Cabrera:2013} further extended the model
to include the non-universality in gaugino and Higgs masses, and found
qualitatively similar results.  The most probable masses of neutralino dark
matter were found to be around 1~TeV, for higgsino dark matter, and 3~TeV, for
wino dark matter.

The wino dark matter case is already in some tension with searches for
gamma-ray lines~\cite{Cohen:2013, Fan:2013, Hryczuk:2014}.  The wino dark
matter annihilation cross-section is significantly larger than the canonical
value for WIMPs due to a non-perturbative effect known as Sommerfeld
Enhancement (SE)~\cite{Hisano:2003ec, Hisano:2004ds, Cirelli:2007xd,
ArkaniHamed:2008qn, Hryczuk:2011}.  The SE calculation for heavy WIMPs is
subject to large logarithmic corrections due to the large hierarchy between the
dark matter mass and the $W$-boson mass.  Reference~\cite{Hryczuk:2012} showed
that full one-loop computation makes the cross-section smaller up to about 30\%
with respect to the SE correction at tree level. Even more precise computations
using Soft Collinear Effective Theory were recently
presented~\cite{Baumgart:2014vma, Bauer:2014ula, Ovanesyan:2014fwa,
Baumgart:2014saa}. Reference~\cite{Baumgart:2014saa} found that when
calculating leading-log semi-inclusive rates, the effect of higher order
corrections are very modest.

There are several studies in the literature that have incorporated the impact
of the Higgs mass measurement, and of XENON-100~\cite{XENON100} and
LUX~\cite{LUX} bounds, on the WIMP-nucleon spin-independent scattering
cross-section on constrained MSSM scenarios (see,
e.g.,~Refs.~\cite{Cabrera:2012, Silverwood:2012tp, Akula:2012,
Buchmueller:2012, Fowlie:2012, Cabrera:2013, Buchmueller:2013rsa,
Roszkowski:2014wqa}). Various statistical approaches have been used to infer
the most probable regions of these scenarios. The parameter space is often
restricted to energies below a few TeV, according to what is expected from
``natural'' supersymmetry.  Interestingly, when performing a proper Bayesian
analysis, the fine-tuning penalization arises automatically from very basic
statistical arguments (the Bayesian version of ``Occam's razor''), allowing to
explore larger regions of the parameter space while taking the notion of
naturalness automatically into account (see, e.g.,~Ref.~\cite{Cabrera:2008tj}).

Reference~\cite{Cabrera:2013} studied the non-universal higgsino and gaugino
masses model within the Bayesian framework. The authors showed that the most
probable regions for the neutralino dark matter are around 1 TeV (higgsino dark
matter) and 3 TeV (wino dark matter), the high masses being mostly due to the
Higgs mass and the relic density constraint.  Although this leaves a very large
portion of the favoured supersymmetric parameter space outside the reach of
LHC, prospects for future dark matter experiments were shown to be very
promising, in particular for XENON-1T~\cite{Aprile:2012} for direct detection,
and for the Cherenkov Telescope Array (CTA)~\cite{Actis2011} for indirect
searches.  Recently, Ref.~\cite{Roszkowski:2014iqa} presented a first study of
the detection prospects for CTA and XENON-1T for neutralino dark matter in the
19-parameter phenomenological MSSM. In this study, the SE was effectively
included as an extrapolation of the results from Refs.~\cite{Hryczuk:2011,
Cirelli:2007xd}.

In the present paper, we study consequences of MSSM models with extended nine
parameters (MSSM-9), and prospects for indirect and direct dark matter
searches.  In addition to the non-universality of the gaugino and Higgs masses
studied in Ref.~\cite{Cabrera:2013}, we investigate the non-universality of
masses and the trilinear couplings in the sfermion sector (i.e., they are
independent between sleptons and squarks).  This treatment is more general
because constraints from the LHC mainly affect the squark and gluino sector and
do not directly reflect on the sleptons.  Since the preferred dark matter
masses are at 1~TeV and above, we discuss constraints from the \emph{Fermi}
gamma-ray satellite and current generation of Cherenkov telescopes, in
particular HESS~\cite{HESS}.  As mentioned above, the wino dark matter around
3~TeV is subject to the SE correction of the annihilation
cross-section~\cite{Hisano:2002fk, Hisano:2004ds, Hryczuk:2011, Hryczuk:2012},
and, therefore, we calculate the SE point-by-point in the scan.  In addition,
the annihilation of wino dark matter also yields strong gamma-ray line signals,
which are tightly constrained by the HESS observations of the Galactic
center~\cite{HESS}.  Note that our results are based on a full numerical study
of MSSM parameter scan, and does not rely on a simplifying assumption of {\it
pure} wino case as adopted in Refs.~\cite{Cohen:2013, Fan:2013, Hryczuk:2014}.
We will discuss prospects for future indirect and direct detection experiments,
most notably for CTA that will have an excellent sensitivity for gamma rays
above 100~GeV \cite{Doro:2013, Wood:2013taa, Pierre:2014tra,
Silverwood:2014yza, Lefranc:2015pza}.

The paper is organized as follows.  In section~II, we discuss the model and the
adopted scanning technique. We will summarize the results of the scan in
section~III, and show in section IV prospects for future indirect and direct
detection experiments.  In section V, we give our conclusions.

\section{MSSM models and Higgs mass}

The discovery of the Higgs boson~\cite{LHC-Higgs1, LHC-Higgs2} has completed
the picture of the standard model of particle physics, which has proven an
extremely good description of particle physics up to the TeV scale.  Beyond the
crucial importance of this discovery by itself, this result has far-reaching
consequences for well-motivated candidates of physics beyond the standard
model, such as supersymmetry, and in particular for the MSSM.

The rather high reported Higgs mass $m_h$ shifts the scale of supersymmetry to
higher values.  In the MSSM, the tree-level Higgs mass is bounded by the mass
of the $Z$-boson, and, therefore, large radiative corrections are needed in
order to reconcile theory and experiment.  An approximate analytic formula for
$m_h$ \cite{Carena:1995bx, Haber:1996fp} reads
\begin{eqnarray}
    \label{mhaprox}
    m_h^2 &\simeq& M_Z^2\cos^2 2\beta +
    \frac{3}{4\pi^2}\frac{m_t^4}{v^2}
    \left[\log\frac{M^2_{\rm SUSY} }{m^2_t}
        \right.\nonumber\\&&{}\left.
        +\frac{X_t^2}{M^2_{\rm SUSY}} \left (1-
            \frac{X_t^2}{12 M^2_{\rm SUSY}}
        \right)
    \right]+\cdots ,
\end{eqnarray}
where $\tan\beta$ is the ratio between the vacuum expectation values of the two
Higgs doublets $v_u=\ev{H_u^0}$ and $v_d=\ev{H_d^0}$, $v^2=v_u^2+v_d^2$, $m_t$
is the top running mass, and $M_{\rm SUSY}$ represents a certain average of
stop masses.  $X_t=A_t-\mu\cot{\beta}$, where $\mu$ is the Higgs mass term in
the superpotential, and $A_t$ is the trilinear stop coupling, both at the
electroweak breaking scale. The first term of Eq.~(\ref{mhaprox}) is the
tree-level Higgs mass, while the second two terms are the dominant radiative
and threshold corrections.  Note that the radiative corrections grow
logarithmically with the stop masses while the threshold corrections has a
maximum for $X_t = \pm \sqrt{6} M_{\rm SUSY}$.  To achieve $m_h \simeq 126$
GeV, one typically needs stop masses larger than $\sim$3~TeV, unless $X_t$ is
close to its maximum value.

In order to evaluate the sensitivity on dark matter in a more generic context,
we parameterize the MSSM with 10 fundamental parameters at the gauge coupling
unification scale. After requiring the correct electroweak symmetry breaking,
we end up with 9 effective parameters:
\begin{eqnarray}
  \left\{ s, M_1, M_2, M_3, m_0^{\tilde q}, m_0^{\tilde l}, m_H, A_0^{\tilde q},
   A_0^{\tilde l}, \tan{\beta}, {\rm sgn}(\mu) \right\},
\end{eqnarray}
where $s$ represents the SM nuisance parameters, $M_1, M_2, M_3$ are the
gaugino masses, $m_0^{\tilde q}$, $m_0^{\tilde l}$, $m_{H}$ are the squark,
slepton, and Higgs masses ($m_H=m_{H_u}=m_{H_d}$), and $A_0^{\tilde q}$ and
$A_0^{\tilde l}$ are the squark and slepton trilinear couplings. The sign of
$\mu$ is fixed to $+1$. All the soft parameters defined at gauge
coupling unification scale, except for $\tan{\beta}$ and the SM nuisance
parameters. Compared with Ref.~\cite{Cabrera:2013}, we further generalize the
sfermion sector, by adopting independent values for squarks and sleptons.

\begin{table}
\begin{center}
\begin{tabular}{|l | l l l | l|}
\hline
Observable & Mean value & \multicolumn{2}{c|}{Uncertainties} & Ref. \\
 &   $\mu$      & ${\sigma}$ (exper.)  & $\tau$ (theor.) & \\\hline
$M_W$ [GeV] & 80.399 & 0.023 & 0.015 & \cite{lepwwg} \\
$\sin^2\theta_{eff}$ & 0.23153 & 0.00016 & 0.00015 & \cite{lepwwg} \\
$\mathrm{BR}(\overline{B}\rightarrow X_s\gamma)\times 10^4$ & 3.55
& 0.26 & 0.30 & \cite{hfag}\\ 
$R_{\Delta M_{B_s}}$ & 1.04 & 0.11 & - & \cite{Aaij:2011qx} \\
$\frac{\mathrm{BR}(B_u \rightarrow \tau \nu)}{\mathrm{BR}(B_u \rightarrow \tau
  \nu)_{SM}}$   & 1.63  & 0.54 & - & \cite{hfag}  \\ 
$\Delta_{0-}  \times 10^{2}$   &  3.1 & 2.3  & - & \cite{Aubert:2008af}  \\
$\frac{\mathrm{BR}(B \to D \tau \nu)}{\mathrm{BR}(B \to D e \nu)} \times
10^{2}$ & 41.6 & 12.8 & 3.5  & \cite{Aubert:2007dsa}  \\ 
$R_{l23}$ & 0.999 & 0.007 & -  &  \cite{Antonelli:2008jg}  \\
$\mathrm{BR}(D_s \to \tau \nu) \times 10^{2}$ & 5.38 & 0.32 & 0.2  &
\cite{hfag}  \\ 
$\mathrm{BR}(D_s\to \mu \nu) \times 10^{3}$ & 5.81 & 0.43 & 0.2  & \cite{hfag}
\\ 
$\mathrm{BR}(D \to \mu \nu) \times 10^{4}$  & 3.82  & 0.33 & 0.2  &
\cite{hfag} \\ 
$\Omega_\chi h^2$ & 0.1196 & 0.0031 & 0.012 & \cite{Ade:2013zuv} \\
$m_h$ [GeV] & 125.66  & 0.41  & 2.0 & \cite{moriond2013} \\
$\mathrm{BR}(\overline{B}_s\to\mu^+\mu^-)$ &  $3.2 \times 10^{-9}$ & $1.5
\times 10^{-9}$ & 10\% & \cite{Aaij:2012nna}\\ 
\hline\hline
   &  \multicolumn{3}{c|}{Limit (95\%~$\text{CL }$)} & Ref. \\ \hline 
Sparticle masses  &  \multicolumn{3}{c|}{As in Table~4 of
  Ref.~\cite{deAustri:2006pe}.}  & \\
$m_\chi - \sigma^\text{SI}_{\chi N}$ & \multicolumn{3}{l|}{XENON100 2012
  limits} & \cite{XENON100} \\ 
\hline
\end{tabular}
\end{center}
\caption{Observables used for the computation of the likelihood function
\label{tab:exp_data}}
\end{table}

We perform a Bayesian analysis to generate a map of the relative probability
of different regions of the parameter space. In doing so, the global
likelihood is defined as a multiplication of individual likelihood functions,
where for each quantity we use a gaussian function with mean $\mu$ and
standard deviation $s = \sqrt{\sigma^2+ \tau^2}$, where $\sigma$ is the
experimental uncertainty and $\tau$ represents our estimate of the theoretical
uncertainty. For upper and lower limits we use a gaussian fuction to model the
drop in the likelihood below or above de experimental bound. The explicit form
of the likelihood function is given in ref.~\cite{deAustri:2006pe}, including
in particular a smearing out of experimental errors and limits to include an
appropriate theoretical uncertainty in the observables. We take into
consideration all the available particle physics data described in
table~\ref{tab:exp_data}, including the Z mass, which is effectively included
adding a Jacobian factor\footnote{The Jacobian factor arise after perform a
  change of variable $\{y_i,\mu,B\}\rightarrow \{m_i,M_z,\tan{\beta}\}$ and
  integrating $M_z$ in the posterior probability density function, as
  explained in detail if Ref.\cite{Cabrera:2008tj}.}, electroweak precision
measurements~\cite{LEP}, B-physics observables~\cite{Amhis:2012bh,
  Aaij:2011qx, Aubert:2007dsa, Antonelli:2008jg, Aaij:2012nna}, the Higgs
mass~\cite{LHC-Higgs1,LHC-Higgs2}, and constraints on the WIMP-nucleon
scattering cross-section by XENON-100~\cite{XENON100} and LUX~\cite{LUX}.  In
addition, we assume a scenario with a single dark matter component that is
produced thermally in the early Universe, by including the measured relic
density according to the \emph{Planck} results~\cite{Planck}.  For the relic
density and $\langle\sigma v\rangle$ computation, we take the SE into account
by creating a grid of the enhancement in the $M_2$-$\mu$ plane using the
Hryczuk et~al.~computation method implemented in DarkSE
\cite{Hryczuk:2011,Hryczuk:2012}. For the computation of $\langle\sigma
v\rangle$ in the present day we implemented a function in DarkSE to extract
the enhancement for $v=10^{-3}$ from the Hryczuk \textit{et~al.} computation
(we validated the results with the pure-wino case showed in
\cite{Hryczuk:2012}).

For the priors of the parameters, we adopted both standard and `improved' log
priors (S-log and I-log, respectively), defined in Refs.~\cite{Cabrera:2012,
  Cabrera:2013}, with ranges described in table~\ref{tab:prior_range}.
\begin{table}
  \centering
  \begin{tabular}{|l c l|}
    \hline
    Parameters && Range scanned \\
    \hline
    $M_1,M_2,M_3$ [GeV] && (-$10^6$, $10^6$)\\
    $m_0^{\tilde q}, m_0^{\tilde l}, m_H$ [GeV]&& (10, $10^6$)\\
    $A_0^{\tilde q},A_0^{\tilde l}$ [GeV] && (-$10^6$, $10^6$)\\
    $\tan{\beta}$ && (2, 60)\\ 
    ${\rm sgn}(\mu)$&& +1\\
    \hline
  \end{tabular}
  \caption{Ranges of model parameters adopted in the scan.}
  \label{tab:prior_range}
\end{table}
In the case of the I-log priors, we effectively assume that the parameters are
associated with a common scale (motivated by a common underlying supersymmetry
breaking mechanism). In the following, we show results for the I-log priors
only, but note that those obtained with S-log priors are very similar, since
the data, in particular the relic density constraint and the Higgs mass, turns
out to be very constraining.  For the SM parameters, we used gaussian priors
described in table ~\ref{tab:nuis_params}.
\begin{table}
\begin{center}
\begin{tabular}{|l l l l |}
\hline
 & Gaussian prior  & Range scanned & ref. \\
\hline
$M_t$ [GeV] & $173.2 \pm 0.9$  & (167.0, 178.2) &  \cite{moriond2013} \\
$m_b(m_b)^{\bar{MS}}$ [GeV] & $4.20\pm 0.07$ & (3.92, 4.48) &  \cite{pdg07}\\
$[\alpha_{em}(M_Z)^{\bar{MS}}]^{-1}$ & $127.955 \pm 0.030$ & (127.835, 128.075) &  \cite{pdg07}\\
$\alpha_s(M_Z)^{\bar{MS}}$ & $0.1176 \pm  0.0020$ &  (0.1096, 0.1256) &  \cite{Hagiwara:2006jt}\\
\hline
\end{tabular}
\end{center}
\caption{Nuisance parameters adopted in the scan.}
\label{tab:nuis_params} 
\end{table}
For the numerical analysis, we use the SuperBayeS code~\cite{Strege:2014},
which uses the nested sampling algorithm implemented in
Multinest~\cite{Feroz:2009}, and integrates SoftSusy~\cite{Allanach:2002},
SusyBSG~\cite{Degrassi:2008}, SuperIso~\cite{Mahmoudi:2009},
DarkSusy~\cite{Gondolo:2004}, MicrOMEGAs~\cite{Belanger:2002}, and
DarkSE~\cite{Hryczuk:2011} for the computation of the experimental observable.
The full likelihood function is the product of the individual gaussian
likelihoods associated to each piece of experimental data. In particular, for
Xenon100 we use the likelihood defined in ~\cite{Bertone:2011nj}. The LUX
limit is applied as a step function only for the $2\sigma$ confidence level
points. For a more detailed explanation of the Bayesian analysis relevant for
the results in this paper, we refer the reader to Ref.~\cite{Cabrera:2013}.

\section{Results of the scan}

\begin{figure}[t]
 \begin{center}
     \includegraphics[width=\linewidth]{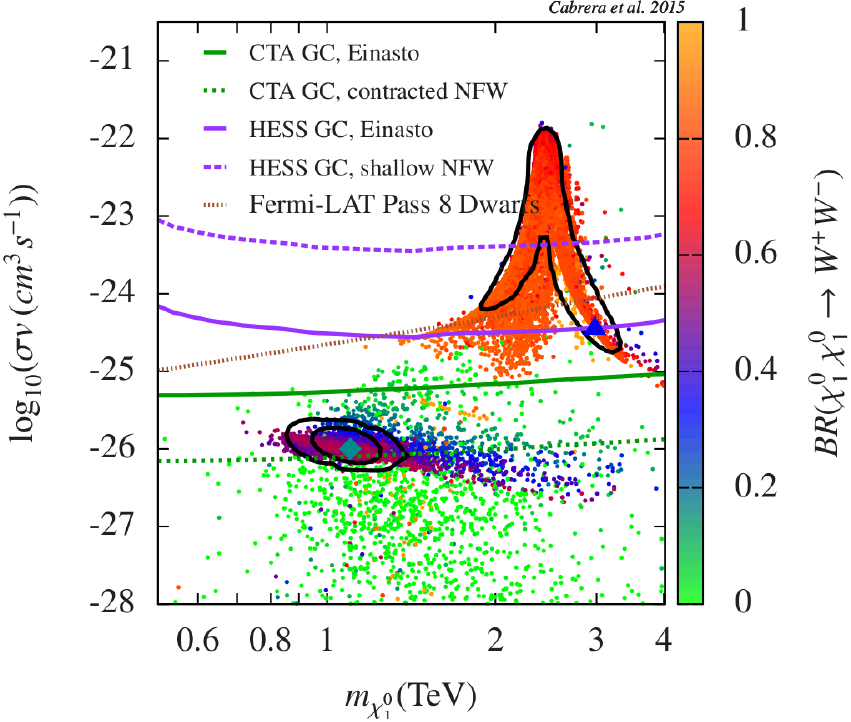}
     \caption{The contours represent 68\% and 95\% posterior probability
       credible regions. Colored points reproduce all the experimental
       observables within $2\sigma$ of confidence level. The cyan diamond
       represent the pure higgsino case from \cite{Profumo:2005xd}, and the
       blue triangle the pure wino case from \cite{Cohen:2013, Hryczuk:2012}.
       The color-coding indicates the branching fraction into $W^+W^-$ final
       states.  The green lines show the sensitivity of CTA as derived in
       Ref.~\cite{Silverwood:2014yza} for $\bar{b}b$ final states and 100h
       observation time (assuming 1\% systematics).  The purple lines show the
       HESS Galactic center limits for generic hadronic final
       states~\cite{Abramowski:2011hc}, adopting an Einasto (solid), a
       contracted NFW (dotted), and a shallow NFW (dashed) profiles The brown
       dotted line shows the \emph{Fermi}-LAT limit from the analysis of the
       dwarf spheroidal galaxies~\cite{FermiDwarfs}. The CTA and
       \emph{Fermi}-LAT limits correspond to $\chi_1^0\chi_1^0\rightarrow b
       \bar{b}$, but limits for $W^+W^-$ final states are very similar.}
  \label{fig:sigmav}
 \end{center}
\end{figure}

\begin{figure}[t]
 \begin{center}
     \includegraphics[width=\linewidth]{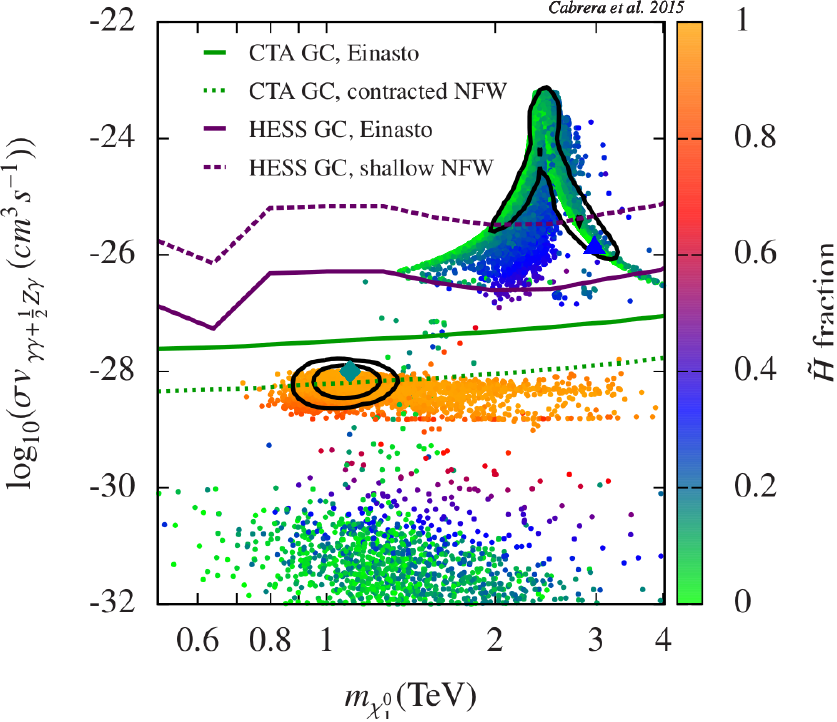}
     \caption{The same as Fig.~\ref{fig:sigmav}, but for annihilation into
     monochromatic photons. The cyan diamond represent the pure higgsino
     case from \cite{Cohen:2013,Hryczuk:2012}, and the blue triangle
     the pure wino case~\cite{Cirelli:2007xd}.  Colors indicate the higgsino
     fraction of the lightest neutralino.  The green lines show the line
     sensitivity of CTA as derived from \cite{Bergstrom:2012vd}, while
     purple lines are the HESS limits \cite{Abramowski:2013ax}, all for the
     Galactic center.}
  \label{fig:sigmav_gamma}
 \end{center}
\end{figure}

In Fig.~\ref{fig:sigmav}, we show two-dimensional contours that represent 68\%
and 95\% credible regions of the most relevant parameters for CTA: the dark
matter neutralino mass and annihilation cross-section.  The posterior has two
peaks in the mass distribution. The largest peak locates around 1 TeV, where
the neutralino mostly consists of higgsino. There is a weaker peak around 3
TeV, where it is mostly wino. The wino dark matter features significantly
larger annihilation cross-section around 10$^{-24}$~cm$^3$~s$^{-1}$ due to SE
correction. Since the SE is a non-relativistic effect causing the distortion of
the wave functions, it is more efficient for heavier particles. We note that
these two regions correspond to those found in Ref.~\cite{Cabrera:2013} with a
seven-parameter MSSM study. In fact, the most probable regions in the posterior
distributions for the mass of the lightest particles (the neutralino) are only
mildly changed compared to \cite{Cabrera:2013}.  This shows the robustness of
the procedure against the number of parameters.

Figure~\ref{fig:sigmav} also shows, as points, regions in the parameter space
that reproduce all experimental observables within $2\sigma$ of confidence
level. We remind that the posterior probability distribution function (PDF)
shows relative probabilities within a model, given the experimental data,
under the hypothesis that the model is correct. The 68$\%$ and 95$\%$
credibility regions show that it is much more likely to find neutralinos with
a mass of $\sim$1 TeV and $\sim$3 TeV, however, these contours not necessarily
cover all the regions that respect the experimental observables. The scattered
points outside the contours show regions that require more tuning to reproduce
the experimental observables and, therefore, their integrated probability is
small.  These less probable regions, with dark matter between 1 TeV and 3 TeV,
correspond to wino-higgsino and wino-bino neutralinos.

An additional region around hundreds GeV, corresponding to bino-like
neutralino, is not show in the figure. Unlike higgsinos and winos, bino
neutralinos can not self-annihilate, therefore, a specific mass relation with
other mass eigenstate is required to have an efficient enough annihilation to
reproduce the correct relic density. For example, a bino quasi-degenerate with
the stau, or a bino mass equal to half of the lightest Higgs or pseudo-scalar
mass. On the other hand, as we mention above, unless we are in the maximal
mixing scenario (which is also subject to certain tuning, see
Ref.~\cite{Casas:2014eca}) the Higgs mass measurement tend to push the spectrum
to higher masses. For few hundreds GeV neutralinos, a fine-tuning is necessary
to reproduce the Higgs mass and the relic density. Therefore, this region has
small statistical weight and is not well explored in our scan. For this reason
we show $m_{\chi_1^0}$ larger that 500 GeV in our figures. In any case, few
hundreds GeV neutralinos will be potentially tested by the LHC.

The annihilation of wino-like dark matter produces relatively strong line
signatures.  Given that the total annihilation cross-section is very large, as
shown in Fig.~\ref{fig:sigmav}, the resulting annihilation cross-section into
$\gamma\gamma$ or $\gamma Z$ channel is correspondingly large.
Figure~\ref{fig:sigmav_gamma} shows the contours of mass and $\langle \sigma
v\rangle_{\gamma\gamma} + \langle \sigma v\rangle_{\gamma Z}/2$, which is the
relevant quantity for line searches with Cherenkov telescopes since the energy
splitting between these two modes are smaller than their typical energy
resolution (e.g., Ref.~\cite{Cohen:2013}).

The colored points in Fig.~\ref{fig:sigmav} show the branching fraction of the
lightest neutralino annihilating into $W^+W^-$.  For the higgsino-like
neutralino ($m_{\chi_1^0}\sim 1$ TeV), the annihilation to $ZZ$ becomes very
important, while for the wino-like neutralino ($m_{\chi_1^0}\sim 3$ TeV), the
annihilation to $W^+W^-$ is the dominant channel. The colored points in
Fig.~\ref{fig:sigmav_gamma} shows the higgsino fraction of the lightest
neutralino.  Note that points inside the 95\% contour are dominantly
higgsino-like and wino-like neutralinos. However, by comparing those points
with the pure higgsino (cyan diamond) and pure wino (blue triangle) cases, it
is clear that the 95\% contour encloses points with small mixing that could
have different characteristics. In particular, in the wino-like region with
$\langle\sigma v\rangle_{\gamma\gamma+\gamma Z/2} > 10^{-27}\rm\;cm^3/s$ and
$m_{\chi_1^0}\sim 2.5$ TeV, the co-annihilation with sfermions plays a very
important role in setting the correct relic density.

\section{Gamma-ray upper limits and prospects}

\begin{figure}[t]
 \begin{center}
     \includegraphics[width=\linewidth]{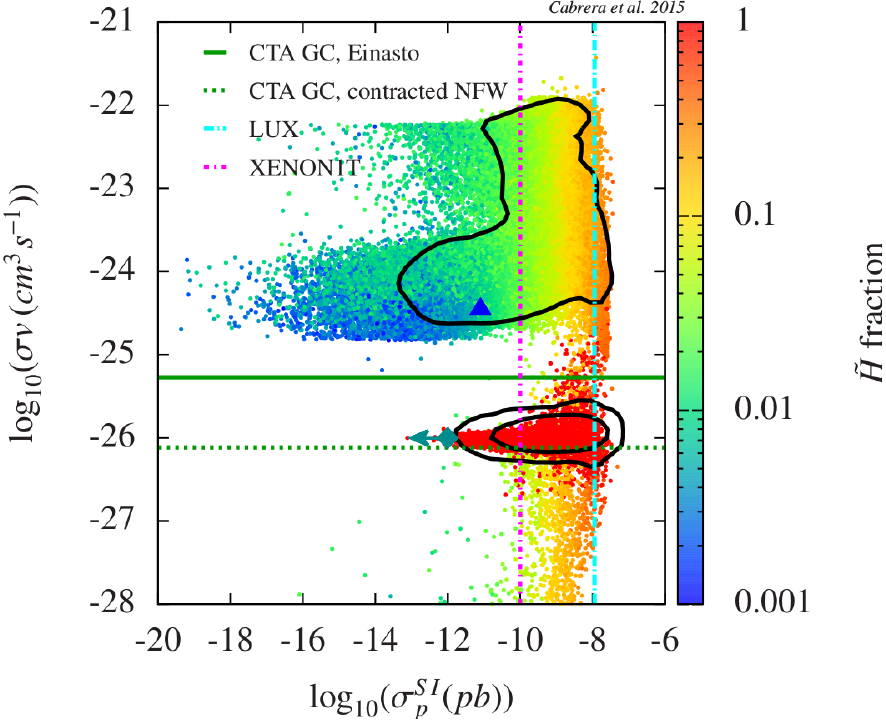}
     \caption{Similar to Fig.~\ref{fig:sigmav}, but showing the total
       annihilation cross-section against the WIMP-proton spin-independent
       cross-section.  The blue triangle and the cyan arrow are the
       theoretical values of the pure wino and pure higgsino case from
       Refs.~\cite{Hill:2013hoa, Hill:2014yxa}. The arrow indicates that the
       value of $\sigma_p^{SI}$ is a theoretical upper limit. The color-coding
       indicates the higgsino fraction.  Green lines show the sensitivity of
       CTA for Einasto (solid) and contracted NFW (dotted) profiles for the
       Galactic center~\cite{Silverwood:2014yza}. Cyan and pink lines
       represent LUX limit~\cite{LUX} and XENON-1T
       sensitivity~\cite{Aprile:2012}. Both the limit and sensitivity lines
       assume $m_{\chi_1^0}=1$~TeV. CTA sensitivity lines correspond to
       $\chi_1^0\chi_1^0\rightarrow b \bar{b}$.}
     \label{fig:scattering}
 \end{center}
\end{figure}

The Galactic center is one of the most promising places to search for signals
from WIMP annihilation, as it is locally the densest region of dark matter,
even though it is often challenging to address astrophysical foregrounds (see,
e.g., Ref.~\cite{Calore2014}). In general, the gamma-ray intensity from
neutralino annihilation towards a direction $\psi$ away from the Galactic
center is given by
\begin{eqnarray}
    I_\gamma(\psi, E_\gamma) &=& \frac{\langle \sigma v \rangle}{8\pi
    m_{\chi_1^0}^2} \frac{dN_{\gamma,{\rm ann}}}{dE_\gamma}
    J(\psi),
    \label{eq:intensity}\\
    \nonumber
    J(\psi) &=& \int_0^{l_{\rm max}} dl \rho_{\chi_1^0}^2 (r[l,\psi]),
\end{eqnarray}
where $dN_{\gamma, {\rm ann}} / dE_\gamma$ represents the annihilation
spectrum, $r(l,\psi)^2 = r_0^2 + l^2 - 2 r_0 l \cos\psi$, $r_0 = 8.5$~kpc is
the Galactocentric radius of the solar system, and $l_{\rm max}$ corresponds to
the virial radius of the Milky Way.  The dark matter density profile in the
Galactic halo $\rho_{\chi_1^0}$ is widely assumed to be given by
phenomenological fits to results of dark-matter-only N-body simulations, such
as the Navarro-Frenk-White (NFW) \cite{NFW} or Einasto~\cite{Einasto1,
Einasto2} functions.  We note, however, that these profiles are observationally
confirmed only at relatively large radii, and far less constrained in the inner
regions~\cite{Iocco:2011jz}, where the strongest annihilation signals are
expected. We will comment again on this point below.

\medskip

\paragraph{Current Status.} Figures~\ref{fig:sigmav} and
\ref{fig:sigmav_gamma} show the predicted annihilation cross-section into
continuum photons (dominated by $W^+W^-$, $ZZ$ and $\bar qq$ final states) and
gamma-ray lines, respectively, compared to different experimental limits and
reaches. Our fiducial density profile is given by an Einasto profile (with
parameters $\alpha=0.17$, $\rho_\odot=0.4\,\rm GeV/cm^3$ and $r_s=20\,\rm
kpc$).  The current upper limits on the Galactic center from HESS searches for
gamma rays from $\bar bb$ final states (Fig.~\ref{fig:sigmav})
\cite{Abramowski:2011hc} \cite{Abramowski:2013ax} and for gamma-ray lines
(Fig.~\ref{fig:sigmav_gamma}) \cite{Abramowski:2013ax} are already very
tight. We find that the wino dark matter region around 3~TeV is almost
completely excluded by the HESS upper limits. This is in agreement with the
findings of Refs.~\cite{Cohen:2013, Fan:2013, Hryczuk:2014}, but interpolating
in the $\sigma v$ enhancement from a $M_2--\mu$ grid instead of extrapolated
from existing pure wino calculations as is in Ref.~\cite{Roszkowski:2014iqa}.

We note, however, that upper limits are still subject to uncertainties mainly
related to the density profile~\cite{Cohen:2013}.  While state-of-the-art
N-body simulations prefer either NFW or Einasto profiles, baryonic effects can
potentially modify the density profiles significantly. For example, baryonic
adiabatic contraction can compress the dark matter profiles, while supernova
feedback can cause the profiles to be shallower (see, e.g.,
Refs.~\cite{Prada2004,Pontzen2012,GomezVargas2013,DelPopolo2015}).  To
illustrate this point, we show in Figs.~\ref{fig:sigmav} and
\ref{fig:sigmav_gamma} how the HESS limits weaken when a shallower dark matter
profile is adopted.  To this end, we use a generalized NFW profile with an
inner slope of $\gamma = 0.7$ (and $r_s = 20\,\rm kpc$, $\rho_\odot = 0.4\,\rm
GeV/cm^3$), which is still in agreement with kinematic and microlensing
observations~\cite{Iocco:2011jz}. In this case, the limits indeed are weakened
and part of the wino best-fit region is still allowed.  A similar effect will
occur for cored profiles. 

In Fig.~\ref{fig:sigmav}, we also show the \emph{Fermi}-LAT limits
from the observation of dwarf spheroidal galaxies from
Ref.~\cite{FermiDwarfs},
which already include the uncertainties in the dark
matter profile and can be hence considered robust (i.e., this represent the
upper end of the uncertainty band).  They exclude most of the wino parameter
space.

\medskip

\paragraph{Prospects for CTA.} Current constraints leave the 1-TeV higgsino
dark matter as most interesting dark matter candidate in the MSSM-9.  The CTA
sensitivities for both the total annihilation cross
section~\cite{Silverwood:2014yza} and for the gamma-ray
lines~\cite{Bergstrom:2012vd} are shown in Figs.~\ref{fig:sigmav} and
\ref{fig:sigmav_gamma}, respectively. These figures show that, for standard
Einasto profiles, it will be challenging for CTA to reach the 1-TeV higgsino
parameter space, unless background systematics are under control at the
sub-percent level~\cite{Silverwood:2014yza}.  However, as explained above,
baryonic effects could potentially increase the chances for a CTA discovery of
higgsino dark matter as baryons can drag dark matter towards the Galactic
center during their cooling, leading to a more cuspy profile~\cite{Prada2004}.
To illustrate this effect, we additionally show in Fig.~\ref{fig:sigmav} the
reach of CTA when a slightly contracted NFW profile, with an inner slope of
$\gamma=1.3$ (and otherwise parameters as above), is adopted. In this case, CTA
has the potential to rule out (or discover) a large part of the best-fit
higgsino dark matter region. Our results are less stringent than the
results found in Ref.~\cite{Roszkowski:2014iqa}, since our estimates for the
CTA sensitivity also include an estimated 1\% systematic uncertainty.  Note also
that a more accurate treatment of the cosmic ray background in
Ref.~\cite{Lefranc:2015pza} leads, for dark matter masses above 1~TeV, to
slightly less stringent projected limits than what we show here, within a
factor of two.

\medskip

\paragraph{Direct Detection.} Fig.~\ref{fig:scattering} shows the most
probable regions plotted for the annihilation cross-section and the
spin-independent scattering cross-section $\sigma_p^{\it SI}$ at tree
level. To give a reference of the size of $\sigma_p^{\it SI}$ for pure cases,
we also show higgsino neutralino (cyan diamond) and wino neutralino (blue
triangle) one-loop computation performed by
\cite{Hill:2013hoa,Hill:2014yxa}.\footnote{The tree level scattering of the
  neutralino with the nucleon, through the Higgs boson, requires a neutralino
  with non-negligible higgsino-bino or higgsino-wino mixing. The other
  possibility is the scattering via squarks, in that case squarks should be
  light enough to give a sizable contribution.}  In the pure higgsino case,
the perturbative QCD and hadronic input $1\sigma$-uncertainties allow only to
set a maximum value for $\sigma_p^{\it SI}$, which is represented in the
figure by an arrow.\footnote{For the computation of $\sigma^{SI}$ we have used
  the s-quarks nucleon form factor derived from measurements of the
  pion-nucleon sigma term \cite{Ellis:2008hf}. On the other hand,
  refs.~\cite{Hill:2013hoa,Hill:2014yxa} use the lattice calculation value.
  Using the lattice value the tree level cross section drops by a factor of ∼
  4 ref~\cite{Feng:2011aa}.} Hence, the mixing of the neutralino has a crucial
role for the computation of $\sigma_p^{\it SI}$. The colored points in
Fig.~\ref{fig:scattering} represent the higgsino composition of the lightest
neutralino, and show how the value of $\sigma_p^{\it SI}$ decreases with the
higgsino fraction in the wino-like region. We remind the reader that this
computation was performed at tree level in our scan. However, almost pure wino
points appear in the region where $\sigma^{SI}$ is smaller that $\sim
10^{-11}$~pb and $\sigma v$ larger than $10^{-25}$~cm$^3$~s$^{-1}$ in
Fig.~\ref{fig:scattering}. Comparing these $\sigma^{SI}$ values with the one
of the pure wino case (blue triangle), it is clear that the one-loop
contribution is the dominant one. Therefore, we would expect that after
including higher order corrections those points will get $\sigma^{SI}$ values
of $\sim 10^{-11}$ pb.

Figure~\ref{fig:scattering} also shows the sensitivities of
XENON-1T~\cite{Aprile:2012} and CTA~\cite{Silverwood:2014yza} (both for
$m_{\chi_1^0} = 1$~TeV), showing that both direct and indirect searches are
very important for the potential discovery of TeV dark matter, that is, at the
moment, the most probable solution in the context of MSSM-9. Note also that the
region around the almost-pure and pure wino and higgsino neutralinos will be
probed by CTA only.

\section{Conclusions}

We studied the prospects for indirect and direct dark matter searches in
context of the MSSM-9 by means of a Bayesian Monte Carlo scan. We find as the
two most likely regions the 1-TeV higgsinos and 3-TeV winos dark matter.
Current limits from dwarf spheroidal observations as well as observations of
the Galactic center with \emph{Fermi}-LAT and HESS exclude almost all the
models with wino-like dark matter, even for flattened profiles of the dark
matter halo.  However, models with 1-TeV higgsino-like dark matter remain
unconstrained.  We find that for regular dark matter profiles, it will be
challenging for CTA to probe the higgsino dark matter parameter space, both for
continuum and gamma-ray line searches.  However, a mildly contracted profile
would improve the prospects significantly, and make most of the higgsino dark
matter parameter space testable in the upcoming years, providing complementary
constraints to future direct detection experiments like XENON-1T.

\acknowledgments

We thank Andrzej Hryczuk for providing us numerical routines and giving useful
advices for the calculation of the Sommerfeld enhancement, Paolo Panci and
Mattia Fornasa for useful discussions.  Furthermore, we thank the anonymous
referee for pointing out an error in the production of $\bar{f}f$ final states
in the first version of this manuscript.
This research was supported by the
Munich Institute for Astro- and Particle Physics (MIAPP) of the DFG cluster of
excellence ``Origin and Structure of the Universe''.  Furthermore, the work
was supported by NWO through one Veni and two Vidi grants (SA, CW and FZ), 
by Funda\c{c}\~ao de Amparo \`a Pesquisa do Estado de S\~ao Paulo (MECC) and
by the Spanish MICINNs Consolider-Ingenio2010 Programme under grant MultiDark
CSD2009-00064, AYA10-21231.

\bibliographystyle{apsrev4-1}
\bibliography{references}

\end{document}